\documentclass[12pt]{article}
\usepackage{amsmath}
\usepackage{amssymb}
\usepackage{cite}
\usepackage{array}
\usepackage{a4}
\usepackage{textcomp}

\begin{document}
\title{Quantum corrections to the mass of the supersymmetric vortex}
\author{D. V. Vassilevich\thanks{Also at V.~A.~Fock Institute of Physics,
St.~Petersburg University, 198904 St.~Petersburg, Russia; e.mail:
Dmitri.Vassilevich@itp.uni-leipzig.de}\\
{\it Institut f\"{u}r Theoretische Physik,
Univerist\"{a}t Leipzig,}\\{\it  Augustusplatz 10, 04109 Leipzig, Germany}}
\maketitle
\begin{abstract}
We calculate quantum corrections to the mass of the vortex in $N=2$
supersymmetric abelian Higgs model in $2+1$ dimensions. We put the system
in a box and apply the zeta function regularization. The boundary conditions 
inevitably violate a part of the supersymmetries. Remaining supersymmetry
is however enough to ensure isospectrality of relevant operators in bosonic
and fermionic sectors. A non-zero correction to the mass of the vortex
comes from finite renormalization of couplings. 
\end{abstract}
\section{Introduction}
The Abrikosov-Nielsen-Olesen (ANO) vortices \cite{ano} play an important
role in modern particle physics \cite{Achucarro:1999it}.
In particular, supersymmetric ANO vortices are essential for understanding
of the monopole condensation (see, e.g., \cite{Vainshtein:2000hu} and
references therein). In $2+1$ dimensions the relation between extended
$N=2$ supersymmetry and the BPS bound has been demonstrated in
\cite{Edelstein:bb} following a more general 
discussion of \cite{Hlousek:1991ej}. 

Quantum corrections to the mass of the supersymmetric ANO vortex\footnote{
Since no analytic form for the profile functions of the ANO vortex is
available, calculations of the mass shift in a non-supersymmetric case
is a rather complicated problem. Recently the fermionic contribution
to the vacuum energy was calculated in a toy model closely resembling 
the abelian Higgs model \cite{BorDro}.} in $2+1$
dimensions were calculated in \cite{Schmidt:cu}, the Chern-Simons terms
were included in \cite{Lee:1994pm}. Both papers \cite{Schmidt:cu,Lee:1994pm}
give zero result for the mass shift. The authors used the arguments 
similar to that of Imbimbo and Mukhi \cite{Imbimbo:1984nq}\footnote{Note, 
that the authors \cite{Imbimbo:1984nq}
used these arguments to show saturation of the
Bogomolny bound (i.e. to estimate the difference between quantum corrections
to the mass and to the central charge) rather than to calculate the mass shift
itself.}
based on the non-local index
theorem of \cite{Callias:1977kg}
and its generalisation by E.~Weinberg \cite{Weinberg:er}.
Roughly speaking, the line of reasoning in \cite{Schmidt:cu,Lee:1994pm}
was as follows. The index theorem was used to show that
\begin{equation}
\rho_B(\omega ) - \rho_F(\omega )\propto \delta (\omega ) \,,
\label{rrd}
\end{equation}
where $\rho_{B,F}$ are the spectral densities in the bosonic and fermionic
sectors respectively. Then the mass shift was identified as
\begin{equation}
\Delta E\propto \int d\omega\, \omega (\rho_B(\omega ) - \rho_F(\omega ))
\,.\label{DEspectdens}
\end{equation}
Due to (\ref{rrd}) the mass shift (\ref{DEspectdens}) should be zero.
In this way, the authors \cite{Schmidt:cu,Lee:1994pm} avoided explicit use
of any regularization. There is however a loophole in this kind of
arguments. First of all, it is assumed that there is a regularization
which supports the mode-by-mode cancellations needed to apply 
(\ref{DEspectdens}). Such regularizations are indeed available.
One of them is the zeta function regularization \cite{zfun}.
However, it requires transition to the discrete spectrum
at least at intermediate
steps. In other words, one has to put the system in a box and
impose some boundary conditions. There is no a priori guarantee that
these boundary conditions can be chosen in such a way to preserve
the index theorem arguments. Besides, without regularising the whole
theory with arbitrary, not only BPS background fields, one cannot
control finite renormalization of charges which are present in the
model.

Quantum corrections to $2+1$ dimensional solitons
should have been re-considered already some time ago. Recent
years have seen a considerable increase of interest to quantum
effects around supersymmetric solitons in $1+1$ dimensions
initiated by the papers \cite{skink} resulted in some very interesting
developments in this field (see \cite{Wimmer:2001yn} for a literature
survey). 

In this paper we re-calculate quantum corrections to the mass of the
supersymmetric vortex using the method \cite{Bordag:2002dg}
applied previously to the supersymmetric kink.
We put the vortex in a box with a circular boundary and impose the
boundary condition which preserve as many symmetries as possible.
We find that one half of the supersymmetries of the vortex is
inevitably broken at the boundary\footnote{BPS states preserve a
half of the supersymmetries of the theory. Since boundaries break
another half, we have a quarter of the original $N=2$
supersymmetry.}. This is however, enough to ensure coincidence of
the eigenfrequencies of the bosonic and fermionic fluctuations. We
then conclude that the total energy of the vortex and the
boundaries is zero. At the next step, we define the energy
associated with the boundaries and find that it is also zero.
Therefore, the whole mass shift of the vortex is due to the finite
renormalization of couplings\footnote{I am gratefull to R.~Wimmer
for pointing out imporatnce of the finite renormalization effects.}. 
It is not zero and is given by the equation
(\ref{finres}) below.

Formally the zero point energy can be represented as
\begin{equation}
\Delta E=\Delta E_B - \Delta E_F,\qquad \Delta E_{B,F} = \frac 12
\sum_{\omega_{B,F}} \omega_{B,F} \,,\label{DEBF}
\end{equation}
where $\omega_{B,F}$ are eigenfrequencies of bosonic and fermionic
fluctuations. The sums in (\ref{DEBF}) are divergent and must be
regularized. We use the zeta function regularization \cite{zfun}:
\begin{equation}
\Delta E^{\rm reg}_{B,F}=\frac 12 \sum_{\omega_{B,F}\ne 0} \omega_{B,F}^{1-2s}
\,,
\label{DEreg}
\end{equation}
where $s$ is the regularization parameter. Note, that zero frequencies
(which do not contribute to (\ref{DEBF}) anyhow) should be explicitly 
excluded.

This paper is organised as follows. 
In the next section we describe properties of classical solutions
and define the operators acting on quantum fluctuation in the
abelian Higgs model without boundaries. In sec.\ \ref{bcsec} we
define gauge invariant boundary conditions which ensure that all
non-zero eigenfrequencies in fermionic and bosonic sectors
coincide. In sec.\ \ref{ssbsec} we analyse supersymmetry of these
boundary conditions and find that one half of the
superinvariancies of the vortex is broken. Section \ref{masssec} is
devoted to calculation of the mass shift. Some concluding remarks
are given in sec.\ \ref{conclu}. Technical details of the
calculations related to the boundary supersymmetries are presented
in the Appendix.

%%%%%%%%%%%%
\section{The model}
This section is devoted to some known properties of the supersymmetric vortices
on manifolds without boundaries. Here we mostly follow 
\cite{Lee:yc,Lee:1994pm}.
%%%%%%%
\subsection{Classical theory}
The Lagrangian of $N=2$
supersymmetric abelian Higgs model in $(2+1)$ dimensions reads:
\begin{eqnarray} 
&&{\cal L}={\cal L}_B+{\cal L}_F \,,\label{LLL}\\
&&{\cal L}_B=-\frac 14 F_{\mu\nu}F^{\mu\nu} -|D_\mu \phi |^2
             -\frac 12 (\partial_\mu w )^2 
             -\frac{e^2}2 \left( |\phi|^2 -v^2 \right)^2
             -e^2 w^2 |\phi|^2 \,,\label{LB} \\
&&{\cal L}_F=i\bar\psi \gamma^\mu D_\mu \psi 
             +i\bar\chi \gamma^\mu \partial_\mu \chi
             -i \sqrt{2} e (\bar\psi \chi \phi -\bar\chi\psi \phi^*)
             +ew\bar\psi\psi \,,\label{LF}
\end{eqnarray}
where $w$  (respectively, $\phi$) is real (respectively, complex) scalar, 
$\psi$ and $\chi$
are two-component complex spinors. $v$ is a constant. 
The signature of the metric $g^{\mu\nu}$ is $(-++)$. As usual, 
$F_{\mu\nu}=\partial_\mu A_\nu -\partial_\nu A_\mu$ is the field strength.
$D_\mu$ is gauge covariant derivative, 
$D_\mu \phi =(\partial_\mu -ieA_\mu )\phi$. The action (\ref{LLL}) is invariant
under the supersymmetry transformations
\begin{eqnarray}
&&\delta A_\mu =i\left( \bar\eta \gamma_\mu \chi -
\bar\chi\gamma_\mu \eta \right) \,,\nonumber \\
&&\delta \phi = \sqrt{2} \bar\eta \psi\,,\qquad
\delta w =i (\bar\chi \eta -\bar \eta \chi ) \,,\nonumber \\
&&\delta \chi = \gamma^\mu \eta \left( \partial_\mu w +
\frac i2 \epsilon_{\mu\nu\lambda}F^{\nu\lambda} \right)
+i\eta (e |\phi |^2 -ev^2 ) \,,\nonumber \\
&&\delta \psi =-\sqrt{2} \left( i\gamma^\mu \eta D_\mu \phi
-\eta e w \phi \right) \label{susy}
\end{eqnarray}
with complex constant spinor parameter $\eta$. $\epsilon^{\mu\nu\rho}$
is the Levi-Civita tensor, $\epsilon^{012}=1$.
The gamma matrices
\begin{equation}
\gamma^0=\left( \begin{array}{cc}
1&0\\
0&-1 \end{array}\right)\,,\qquad
\gamma^1=\left( \begin{array}{cc}
0&1\\
-1&0 \end{array}\right)\,,\qquad
\gamma^2=\left( \begin{array}{cc}
0&i\\
i&0 \end{array}\right) \label{gamma}
\end{equation}
satisfy the equation
\begin{equation}
\gamma^\mu \gamma^\nu =-g^{\mu\nu} -i\epsilon^{\mu\nu\rho}\gamma_\rho
\,.\label{gagag}
\end{equation}
We shall mark upper (lower) components of all spinors with the
subscript ``$+$'' (respectively, ``$-$''), so that
\begin{equation}
\eta=\left( \begin{array}{c} \eta_+ \\ \eta_- \end{array} \right),
\label{spinorcom}
\end{equation}
for example.

Consider now static bosonic field configurations such that $A_0=w=0$.
Such configurations are invariant with respect to one half of
the supersymmetry
transformations (\ref{susy}) corresponding to $\eta_+=0$ if and only
if
\begin{eqnarray}
&&(D_1+iD_2)\phi =0\,,\label{bog1}\\
&&F_{12}+e(|\phi |^2 -v^2 )=0 \,.\label{bog2}
\end{eqnarray}
These are just the Bogomolny \cite{Bogomolny:1975de}
self-duality equations.

The classical vortices 
\begin{equation}
\phi =f(r) e^{in\theta},\qquad eA_j=\epsilon_{jk} \frac {x^k}{r^2}
(a(r)-n) \label{vortex}
\end{equation}
satisfy (\ref{bog1}) and (\ref{bog2}) if 
\begin{eqnarray}
&&\frac 1r \frac{d}{dr} a(r)=e^2\left( f^2(r)-v^2 \right)\,,\nonumber\\
&&r \frac{d}{dr} \ln f(r) = a(r) \,.\label{af-eqs}
\end{eqnarray}
In these equations $n\in \mathbb{N}$ is vorticity 
(which we assume to be positive), $j,k\in \{1,2\}$, $\epsilon_{12}=1$,
and $r$, $\theta$ are usual polar coordinates on the plane. 
The functions $f(r)$ and $a(r)$ satisfy the conditions
\begin{eqnarray}
&& f(0)=0\,,\qquad f(\infty )=v \,,\label{cononf}\\
&& a(0)=n\,,\qquad a(\infty )=0 \,.\label{conona}
\end{eqnarray}
The classical energy of this configuration reads 
(see, e.g. \cite{Lee:1994pm}):
\begin{equation}
E^{\rm cl}=2\pi n v^2 \,.\label{Eclass}
\end{equation}
%%%%%%%%
\subsection{Quantum fluctuations}
Let us now turn to quantum fluctuations about the background
(\ref{vortex}). We shift $\phi \to \phi +\varphi$ and
$A_\mu \to A_\mu +\alpha_\mu$, where $\varphi$ and $\alpha_\mu$ are
the fluctuations. Since all other fields are zero on the background,
we do not need to introduce more notations. It is convenient
to use the background gauge fixing term\footnote{This gauge condition
belongs to the so-called $R_\xi$ family 
\cite{Rxigauge}.}
\begin{equation}
{\cal L}_{\rm gf}=-\frac 12 \left[ \partial_\mu \alpha^\mu 
-ie (\varphi^* \phi - \varphi \phi^*) \right]^2\,,\label{gfterm}
\end{equation}
which generates the following action for the complex ghosts $\sigma$:
\begin{equation}
{\cal L}_{\rm ghost}= \sigma^* ( \partial_\mu \partial^\mu 
-2e^2 \phi \phi^* )\sigma \,.\label{ghostac}
\end{equation}

Next we expand the action (\ref{LB}) about the classical background.
The terms linear in fluctuations vanish due to the equations of
motion. In the next, quadratic, order we have in the bosonic sector:
\begin{eqnarray}
&&{\cal L}_B^2 + {\cal L}_{\rm gf}=-\frac 12 \alpha_\mu 
(\Box -2e^2\phi^*\phi) \alpha^\mu
-(D_\mu \varphi )(D^\mu \varphi)^* -e^2 \varphi \varphi^*
(3\phi\phi^* -v^2) \nonumber\\
&& \qquad\qquad -2ie \alpha^\mu \left( \varphi^* D_\mu \phi -
\varphi D_\mu \phi^* \right) -\frac 12 (\partial_\mu w )^2 
-e^2 w^2 |\phi|^2 \,,\label{L2bos}
\end{eqnarray}
where the covariant derivative $D_\mu$ depends on 
on the background gauge potential $A_\mu$;
$\Box =\partial_\mu \partial^\mu$.

The quadratic part of the fermionic action coincides with
(\ref{LF}) where all bosonic fields take their background values
(so that $w=0$, for example). Therefore, the equation which defines
eigenfrequencies $\omega_F$ in the fermionic sector reads:
\begin{equation}
\omega_F \left( \begin{array}{c} \psi \\ \chi \end{array} \right)
:=i\partial_0 
\left( \begin{array}{c} \psi \\ \chi \end{array} \right)
=-i\gamma^0 
\left( \begin{array}{cc} \gamma^j D_j ,& -\sqrt{2} e\phi \\
                         \sqrt{2} e\phi^* ,& \gamma^k \partial_k
\end{array} \right) 
\left( \begin{array}{c} \psi \\ \chi \end{array} \right) \,.
\label{Direq}
\end{equation}
By taking square of this equation one obtains \cite{Lee:1994pm}:
\begin{equation}
\omega_F^2 \left( \begin{array}{c} U \\ V \end{array} \right)
= \left( \begin{array}{cc} D_F D_F^\dag ,& 0 \\
                       0,& D_F^\dag D_F \end{array} \right)
\left( \begin{array}{c} U \\ V \end{array} \right) \,,
\label{sqDir}
\end{equation}
where
\begin{equation}
U=\left( \begin{array}{c} \psi_+ \\ \chi_- \end{array} \right)
\,,\qquad 
V=\left( \begin{array}{c} \psi_- \\ \chi_+ \end{array} \right)
\label{defUV}
\end{equation}
and
\begin{equation}
D_F = \left( \begin{array}{cc} D_+,& -\sqrt{2} e\phi \\
                           -\sqrt{2}e\phi^*,&\partial_-
\end{array}\right)\,,\qquad
-D_F^\dag = \left( \begin{array}{cc} D_-,& \sqrt{2} e\phi \\
                           \sqrt{2}e\phi^*,&\partial_+
\end{array}\right)\,.\label{DFDF}
\end{equation}
In this equation we have used holomorphic and anti-holomorphic
components of two-dimensional differential operators:
\begin{equation}
D_\pm := D_1\pm i D_2\,,\qquad 
\partial_\pm :=\partial_1\pm i \partial_2 \,.\label{Ddpm}
\end{equation}

Let us now return to the bosonic fluctuation. As one can see from
(\ref{L2bos}), the equations for $\alpha_0$ and for $w$ decouple
from the rest of the bosonic modes. Moreover, the squared 
eigenfrequencies of these fields are given by the eigenvalues of
the operator
\begin{equation}
\Delta_w=-\partial_j\partial_j +2e^2|\phi |^2 \,.\label{opwa0}
\end{equation}
The same operator defines also the ghost eigenfrequencies.
Therefore, contributions of all these fields to the vacuum energy
cancel (provided they all satisfy the same boundary conditions).

A very important observation regarding the rest of the bosonic
perturbations was made by Lee and Min \cite{Lee:1994pm}.
They demonstrated that the eigenfrequencies for $\varphi$ and
$\alpha_j$ follow from the equation:
\begin{equation}
\omega_B^2 
\left( \begin{array}{c} \varphi \\ {i\alpha_+}{\sqrt{2}} \end{array} \right)
:= -\partial_0^2
\left( \begin{array}{c} \varphi \\ {i\alpha_+}{\sqrt{2}} \end{array} \right)
= D_F^\dag D 
\left( \begin{array}{c} \varphi \\ {i\alpha_+}{\sqrt{2}} \end{array} \right)\,,
\label{omB}
\end{equation}
where $\alpha_+=\alpha_1+i\alpha_2$.
One can check this statement by a direct calculation using the
Bogomolny equations (\ref{bog1}) and (\ref{bog2}) for the background fields.

For the sake of completeness we give here explicit expressions for
$D_FD_F^\dag$ and $D_F^\dag D_F$:
\begin{eqnarray}
&&D_F^\dag D_F = -\left( \begin{array}{cc} D_j^2 - e^2 (3|\phi |^2
-v^2 ), &  -\sqrt{2} e (D_-\phi ) \\
-\sqrt{2} e (D_+\phi^*), & \partial_j^2 -2e^2 |\phi |^2
\end{array} \right) \label{DdagD}\,,\\
&&D_F D_F^\dag = -\left( \begin{array}{cc}D_j^2 - e^2 (|\phi |^2
+v^2 ), & 0 \\ 0, & \partial_j^2 -2e^2 |\phi |^2
\end{array} \right)\,.\label{DDdag}
\end{eqnarray}
We stress that these formulae are valid only if the background
satisfies the Bogomolny equations.
%%%%%%%%%%%%%%
\section{Boundary conditions}\label{bcsec}
The aim of this section is to define the boundary conditions which
support the factorisation properties of the eigenfrequency equations
(\ref{sqDir}) and (\ref{omB}). We like to keep as much symmetry between
the bosonic and fermionic fluctuation as possible. 

Let us put the system in a spherical 
shell with the boundary at $r=R$ (the time coordinate $x^0$ remains,
of course, unrestricted). The relation
\begin{equation}
u_\pm = e^{\pm i\theta} \left( u_r \pm \frac ir u_\theta \right)\,,
\label{pmrt}
\end{equation}
between complex and angular representations of an arbitrary two-vector $u_j$
will be useful in this and subsequent sections.

We start with gauge invariant boundary conditions for $\alpha_\mu$
and $\sigma$. By gauge invariance we understand the following property
\cite{Vassilevich:1997iz}. Let ${\cal B}^{[\alpha]}$ and 
${\cal B}^{[\sigma]}$ be boundary operators which define boundary conditions
for $\alpha$ and $\sigma$ respectively:
\begin{equation}
{\cal B}^{[\alpha]} \alpha_\mu |_{\partial M}=0,\qquad
{\cal B}^{[\sigma]} \sigma |_{\partial M}=0 \,,\label{bops}
\end{equation}
where $\partial M$ is boundary of the manifold\footnote{For Dirichlet
boundary conditions the operator ${\cal B}$ is just the identity operator,
so that ${\cal B}\phi |_{\partial M}=0$ simply means 
$\phi |_{\partial M}=0$. For Neumann boundary conditions ${\cal B}$
contains a normal derivative ($\partial_r$ in our case). More complicated
boundary operators will be introduced below.}. 
This system is gauge invariant if 
\begin{equation}
{\cal B}^{[\alpha]} \partial_\mu \sigma |_{\partial M}=0 \,.
\label{gauinv}
\end{equation}
This property simply means that space defined by (\ref{bops}) is invariant
under the gauge transformations.

There are only two sets of gauge-invariant local boundary conditions
for the Maxwell field\footnote{This point is discussed in the monographs
\cite{Gilkey:1995,EKP}, see also \cite{Vassilevich:1994cz}.
}. Let us take one of them\footnote{Calculations for the other (dual)
set of the boundary conditions go in a similar manner.}:
\begin{equation}
\alpha_0|_{\partial M}=0,\quad
\alpha_\theta |_{\partial M}=0,\quad
\left( \partial_r +\frac 1r \right) \alpha_r |_{\partial M}=0,\quad
\sigma |_{\partial M}=0.\label{asbc}
\end{equation}
Obviously, if $\sigma$ satisfies Dirichlet boundary conditions,
$\partial_0 \sigma$ and $\partial_\theta \sigma$ also satisfy
Dirichlet boundary conditions since $\partial_0$ and $\partial_\theta$
which act in tangential directions to the boundary. 
A bit more work is needed to show that the
condition for $\alpha_r $ is also gauge invariant. 
Gauge transformation of the boundary condition (\ref{asbc})
for $\alpha_r$ reads:
\begin{equation}
\left( \partial_r +\frac 1r \right) \partial_r \sigma =
\left[ -\Delta_w \sigma \right] +
\left[ - \frac 1{r^2}\partial_\theta^2 + 2\phi\phi^* \right]\sigma
\,,\label{ginar}
\end{equation}
where we added and subtracted several terms such that the first
bracket contains the operator (\ref{opwa0}) which defines eigenfrequencies
in the ghost sector. We can expand $\sigma$ in a sum over eigenfrequencies:
$\sigma =\sum_k \sigma_k$ so that 
$\Delta_w \sigma_k=\omega_k^2\sigma_k$ and each $\sigma_k$
satisfies Dirichlet boundary conditions as required by (\ref{asbc}).
Therefore,
\begin{equation}
\left[ -\Delta_w \sigma \right]|_{\partial M} =
-\sum_k \omega_k^2 \sigma_k |_{\partial M} =0.\label{1stterm}
\end{equation}
This proves that the first term on the right hand side of (\ref{ginar})
vanishes on the boundary. The second term there is also zero on the
boundary since it does not contain normal derivatives acting on 
$\sigma$. We conclude, that the boundary conditions (\ref{asbc})
are indeed gauge invariant. 

Eigenfrequencies of $\sigma$, $\alpha_0$, and $w$ are defined by
the same operator $\Delta_w$. Therefore, it is natural to impose
on $w$ the same (Dirichlet) boundary conditions:
\begin{equation}
w|_{\partial M}=0 \,.\label{bcw}
\end{equation}  

Radial and angular components of $\alpha$ can be expressed through
$\alpha_+$:
\begin{equation}
\alpha_r =\Re \left( e^{-i\theta} \alpha_+ \right)\,,\qquad
\alpha_\theta = r \Im \left( e^{-i\theta }\alpha_+ \right)\,.
\label{rthetaplus}
\end{equation}
The operator $D_F^\dag D_F$ acts on the bosonic fluctuations
$(\varphi ,i\alpha_+/\sqrt{2})$ as well as on the fermionic components
$V$ (cf. (\ref{defUV})). Hence, we impose the same boundary 
conditions on the lower component $V_2=\chi_+$ as we have already
defined for $i\alpha_+ /\sqrt{2}$. Namely,\footnote{\label{ft}Strictly 
speaking, eigenfrequencies of $(\varphi ,i\alpha_+/\sqrt{2})$
and $(V_1,V_2)$ are the same even if we identify respective
boundary conditions up to a common constant phase factor. 
This freedom will be discussed in sec.\ \ref{ssbsec}.}
\begin{equation}
\Re \left( e^{-i\theta} \chi_+ \right) |_{\partial M}=0,
\qquad \left( \partial_r +\frac 1r \right) \Im
\left( e^{-i\theta} \chi_+ \right) |_{\partial M}=0.
\label{bcchip}
\end{equation}

To fix boundary conditions on the rest of the fields we shall use
intertwining relations between $D^\dag_F D_F$ and $D_FD_F^\dag$.
Let $U(\omega )$ and $V(\omega )$ be solutions of (\ref{sqDir})
with $\omega_F=\omega$. We can write formally:
\begin{eqnarray}
&&V(\omega )=\omega^{-2} D_F^\dag U(\omega )\,,\label{VomU}\\
&&U(\omega )=\omega^{-2} D_F V(\omega )\label{UomV}
\end{eqnarray}
for $\omega \ne 0$. We are looking for boundary conditions
compatible with (\ref{VomU}) and (\ref{UomV}). Such boundary 
conditions will ensure that the operators $D^\dag_F D_F$ 
and $D_FD_F^\dag$ have coinciding non-zero eigenvalues.

Let us consider the first line in (\ref{UomV}) which reads:
\begin{equation}
U_1(\omega ) =\omega^{-2} \left( D_+ V_1(\omega ) 
                              -\sqrt{2} e \phi V_2(\omega ) \right).
\label{U1VV}
\end{equation}
Let us suppose that the boundary conditions for all components
$U_1$, $U_2$, $V_1$, $V_2$ are mutually independent. This technical
requirement will simplify the calculations below, but will not affect
our main result. Let us take $V_1=0$ first. Then the first equation
in (\ref{bcchip}) yields:
\begin{equation}
\Re \left( e^{-i\theta}\phi^* U_1 \right)|_{\partial M}=
\Re \left( e^{-i\theta}\phi^* \psi_+ \right)|_{\partial M}=
0\,.\label{bcU1D}
\end{equation}
Note, that we are {\it not} allowed to take normal derivative of
(\ref{U1VV}) {\it after} we have put $V_1=0$ in order to get further
conditions on $U_1$ since $\partial_r^2 V_1$ is related to
$V_2$ by the equations of motion, and, therefore, cannot be
considered as an independent quantity on the boundary.
Instead, we take the other component of (\ref{UomV}):
\begin{equation}
U_2(\omega )=\omega^2 \left( \partial_- V_2 
                            -\sqrt{2} e\phi^* V_1 \right)\,.
\label{U2VV}
\end{equation}
The boundary conditions (\ref{bcchip}) immediately give:
\begin{equation}
\Im \left( U_2 \right)|_{\partial M}=0,\qquad
\Im \left( \phi^* V_1 \right)|_{\partial M}=0 \,.
\label{bcU2V1}
\end{equation}
Next we return to (\ref{U1VV}) and put there $V_2=0$ to
see that
\begin{equation}
\left( \partial_r -2(\partial_r \ln \phi^* ) \right)\Re 
\left( \phi^* V_1 \right)|_{\partial M}=0  \label{bcV1N}
\end{equation}
as a consequence of (\ref{bcU1D}) and (\ref{bcU2V1}). Again, we identify 
the boundary conditions for $V_1$ with those for the first component
of the boson doublet $(\varphi ,i\alpha_+/\sqrt{2})$:
\begin{equation}
\Im \left( \phi^* \varphi \right)|_{\partial M}=0 \,,\qquad
\left( \partial_r -2(\partial_r \ln \phi^* ) \right)\Re 
\left( \phi^* \varphi \right)|_{\partial M}=0 \,.\label{bcphi}
\end{equation}
Similarly, we use (\ref{VomU}) to fix the boundary conditions for
$U_1=\psi_+$ and $U_2=\chi_-$:
\begin{equation}
\partial_r \Re (\chi_-)|_{\partial M}=0\,,\qquad
\left( \partial_r +\frac 1r \right) 
\Im \left( e^{-i\theta} \phi^* \psi_+ \right)|_{\partial M}=0\,.
\label{bcUUN}
\end{equation}

We have found a set of the boundary conditions which guarantees 
coincidence of non-zero eigenfrequencies for bosons and for
fermions. 
We summarise the results of this section in Tables 1 and 2.

\begin{table}
\begin{tabular}{|l|c|c|c|c|}\hline
Field & $\sigma$ & $\alpha$ & $w$ & $\varphi$ 
\\ \hline
Equation & (\ref{asbc})& (\ref{asbc}) & (\ref{bcw}) & (\ref{bcphi}) 
\\ \hline
\end{tabular}\\[2pt] Table 1.
Summary of the boundary conditions for ghosts and bosons. 
\end{table}
\begin{table}
\begin{tabular}{|l|c|c|c|c|}\hline
Field & $\psi_+=U_1$ & $\psi_-=V_1$ & $\chi_+=V_2$ & $\chi_-=U_2$ 
\\ \hline
Equation & (\ref{bcU1D}), (\ref{bcUUN}) & (\ref{bcU2V1}), (\ref{bcV1N}) &
(\ref{bcchip}) & (\ref{bcU2V1}), (\ref{bcUUN})\\ \hline
\end{tabular}\\[2pt]
Table 2. Summary of the boundary conditions for spinors.
\end{table}
%%%%%%%%%%%%%
\section{Supersymmetry breaking at the boundary} \label{ssbsec}
In the previous section we have constructed boundary conditions
which support isospectrality of the operators acting in the
bosonic and fermionic sectors. This suggests that certain degree
of supersymmetry still remains in the problem even in the presence 
of boundaries. Due to the vortex, initial $N=2$ supersymmetry 
(\ref{susy}) is
broken to the transformations with $\eta_+=0$. However, the other
complex component $\eta_-$ of the parameter $\eta$ remains
unrestricted. In this section we show that in the presence of
boundaries supersymmetry is broken to a real subgroup.
 
Let us consider the $\eta_-$ transformation of $\alpha_+$:
\begin{equation}
\delta \alpha_+ =2i\eta_-^* \chi_+ \,.\label{delap}
\end{equation}
From this equation we see that if $\eta_-$ is an arbitrary
{\it complex} parameter, it is not possible to impose 
different supersymmetric boundary conditions on real
and imaginary parts of $\alpha_+$. For example, if 
$\Im (r e^{-i\theta}\alpha_+)=\alpha_\theta$ satisfies
Dirichlet boundary conditions (as in our case), then
because of (\ref{delap})
both real and imaginary parts of $\chi_+$ should also
satisfy Dirichlet boundary conditions. This, in turn,
yields Dirichlet boundary conditions for
$\Re (e^{-i\theta}\alpha_+)=\alpha_r$ contradicting gauge
invariance of the boundary value problem.

However, if we require
\begin{equation}
\Re (\eta_-) =0 \label{Imeta}
\end{equation}
the boundary conditions obtained in the previous section
become invariant under the supersymmetry transformations
of the {\it boson} fields (i.e., boundary conditions are
compatible with first there variations in (\ref{susy})).
This statement can be checked by direct and rather elementary
calculations\footnote{One has to take into account that complex
conjugation of the Grassmann variables also changes order in their
products. For example, $(\eta^*_-\chi_-)^*=\chi_-^*\eta_-$.
Therefore, product of two real Grassmann variables is imaginary.
Forgetting this property one would get $\Im (\eta_-)=0$ instead
of (\ref{Imeta}) and a contradiction with superinvariance of the boundary
conditions for fermions.}. For example, compatibility of (\ref{delap})
is obvious since $\alpha_+$ and $i\chi_+$ satisfy the same
boundary conditions. 

Supertransformations (\ref{susy}) of the fermions are also compatible
with our boundary conditions if $\Re (\eta_-)=0$.
Proof of this statement (which is more involved than in the case of the bosons)
is sketched in Appendix \ref{susyapp}. 

One can change the residual supersymmetry
by using the freedom mentioned above
in the footnote \ref{ft}. Since multiplication by
a constant phase factor commutes with all operators and preserves 
normalisation of the eigenfunctions, one can replace the spinor
field ${\cal F}=(U,V)$ by ${\cal F}_\kappa =e^{i\kappa} {\cal F}$
in the boundary conditions derived in sec.\ \ref{bcsec}. However,
this phase factor can be absorbed in a redefinition of the
supersymmetry transformation parameter: 
$\eta\to\eta_\kappa=e^{i\kappa}\eta$. Then the supersymmetry 
transformations (\ref{susy}) remain the same in terms of 
${\cal F}_\kappa$, $\eta_\kappa$. Supersymmetry of new transformed
boundary condition would therefore require $\Re \eta_{\kappa -}=0$.

Let us stress, that the remaining supersymmetry is enough
 to achieve isospectrality of relevant
operators in the bosonic and fermionic sectors. 
Of course, there is no guarantee that such cancellations
will occur at higher loops as well. To understand the situation
from the non-perturbative point of view one has to modify the
Witten--Olive construction \cite{Witten:mh}  accordingly.

%%%%%%%%%%%%
\section{Quantum corrections to the mass of the vortex}\label{masssec}
In the one-loop approximation the renormalised mass shift of the
vortex consist of three terms:
\begin{equation}
\Delta E^{\rm ren} = \Delta E(V+B)^{\rm ren} - \Delta E (B)^{\rm
ren} + \Delta E^{\rm f.r.} \,,\label{DE3DE}
\end{equation}
where the first term is the zero point energy in for the vortex in
the spherical box, the second term is the energy associated with
the boundaries of the box, and the third term is a contribution
from finite renormalization  of charges in the classical
expression for the mass of the vortex.

In sec.\ \ref{bcsec} we have found such boundary conditions that
all non-zero eigenfrequencies in the bosonic sector coincide with
non-zero eigenfrequencies in the fermionic sector. Therefore, for
a sufficiently large $s$ (cf. eq.\ (\ref{DEreg})),
\begin{equation}
\Delta E_B^{\rm reg} = \frac 12 \sum_{\omega_B\ne 0}
\omega_B^{1-2s} =\frac 12 \sum_{\omega_F\ne 0} \omega_F^{1-2s} =
\Delta E_F^{\rm reg} \,.\label{DEBDEF}
\end{equation}
If now we analytically continue (\ref{DEBDEF}) to $s=0$ we find
that both divergent and finite parts of the vacuum energy for the
vortex in the box are zero,
\begin{equation}
\Delta E(V+B)^{\rm ren} =0 \,.\label{1term}
\end{equation}
This equation is, of course, valid for arbitrary radius $R$ of the
box.

%%%%%
\subsection{Quantum energy of the boundaries}
Here we calculate the vacuum energy of the boundary of the box
in the limit $R\to\infty$. First we have to show that between
some characteristic radius $R_1$ (which is defined essentially by
the size of the vortex) and $R$ the theory may be approximated
by free massive fields.

As $r$ goes to infinity both profile functions of the vortex $f$
and $a$ go exponentially fast to their asymptotic values
(\ref{cononf}), (\ref{conona}). Therefore, near the boundary we
can assume that $a$ and $f$ are constants and neglect their
derivatives. Consequently, the operator $\Delta_w$ which defines
the eigenfrequencies of $w$, $\alpha_0$ and of the ghosts $\sigma$
can be approximated by
\begin{equation}
\tilde \Delta =-\partial_j^2 +2e^2v^2.\label{asop}
\end{equation}

To understand what happens with the rest of the fields as $r\to\infty$
one has to analyse the operators (\ref{DdagD}) and (\ref{DDdag}).
The Bogomolny equation (\ref{bog1}) yields:
\begin{equation}
\partial_r \phi =-\frac ir D_\theta \phi \,.
\end{equation}
Consequently,
\begin{equation}
(D_-\phi )= e^{-i\theta} 2 \partial_r \phi \ {\to}\ 0
\end{equation}
as $r\to \infty$. The same is true for $(D_+\phi^*)$, and both
functions are approaching zero exponentially fast. This means that
for large $r$ the off-diagonal terms in (\ref{DdagD}) can be
neglected. The operators (\ref{DdagD}) and (\ref{DDdag}) contain
the background vector potential (\ref{vortex}) which does not
vanish sufficiently fast at the infinity. This potential can be
however transformed away by the following unitary change of 
variables for charged quantum fluctuations:
\begin{equation}
\tilde \varphi =e^{i\beta(r) \theta} \varphi,\qquad
\tilde \psi =e^{i\beta(r) \theta} \psi,\label{tilfiel}
\end{equation}
where the phase $\beta (r)$ is chosen in such a way that
$\beta (r)=-n$ for $r>R_1$ and $\beta (r)\to 0$ insides the vortex.
It is easy to see that in terms of new fields $\tilde\varphi$ and
$\tilde\psi$ in the asymptotic region the eigenfrequencies are
defined by the free operator (\ref{asop}) up to exponentially
small terms. 

One can easily show that not only the operators, but also the
boundary conditions are identical in the bosonic and fermionic 
sectors up to exponentially small terms. Indeed, the fields
$w$, $\alpha_0$ and $\sigma$ satisfy Dirichlet boundary conditions.
Therefore, their contributions to the vacuum energy cancel
also in the effective theory near the boundary. 
The fields $i\alpha_+$, $\chi_+$, and $\tilde\psi_+$ satisfy:
\begin{eqnarray}
&&\left( \partial_r +\frac 1r \right) \Im 
\left( e^{-i\theta} (i\alpha_+,\chi_+,\tilde\psi_+) \right) 
\vert_{\partial M}=
0\,,\nonumber\\
&&\Re \left( e^{-i\theta} (i\alpha_+,\chi_+,\tilde\psi_+) \right) 
\vert_{\partial M}=0\,.\label{effbc1}
\end{eqnarray}
To derive the effective boundary conditions for $\tilde\psi_+$ we have
used that $e^{-i\beta(r)\theta}\phi^*$ goes exponentially fast
to a constant when $r\to\infty$. Similarly we have:
\begin{eqnarray}
&&\partial_r \Re \left( (\tilde\varphi,\tilde\psi_-,\chi_-) \right)
\vert_{\partial M}=
0\,,\nonumber\\
&&\Im \left( (\tilde\varphi,\tilde\psi_-,\chi_-) \right)
\vert_{\partial M}=0 \,.\label{effbc2}
\end{eqnarray}
Taking into account a relative
factor of $1/2$ in the contributions of spinors to the vacuum
energy, we see that the total quantum energy associated with the
effective field theory near the boundary is zero. This is true for
arbitrary values of the regularization parameter, and, therefore,
\begin{equation}
\Delta E (B)^{\rm ren}=0 \,.\label{DEBren}
\end{equation}

%%%%%
\subsection{Finite renormalization}
As usual the renormalization is performed in the topologically
trivial sector. We put $\phi =const.$ and calculate the effective
potential. We shall not need other background fields. We use again
the zeta function regularization as in (\ref{DEreg}). A real
bosonic field with the mass $m$ contributes to the regularized
effective potential
\begin{equation}
W_m(s)=\frac 12 \sum \omega (m)^{1-2s} = \frac 12 \zeta_m \left(
s-\frac 12\right) \,, \label{mW}
\end{equation}
where $\zeta_m$ is the zeta function for the operator
$\Delta_m=-\partial_j^2+m^2$. It can be expressed through
corresponding heat kernel:
\begin{equation}
\zeta_m \left( s-\frac 12\right)= \Gamma \left( s-\frac
12\right)^{-1} \int d^2x \int\limits_0^\infty dt\, t^{s-\frac 12
-1} K(t,x) \,.\label{zmKt}
\end{equation}
The heat kernel reads
\begin{equation}
K(t,x)=\langle x | e^{-t\Delta_m } | x\rangle = (4\pi t)^{-1}
e^{-m^2t} \,.\label{Ktx}
\end{equation}
The integral over $x$ in (\ref{zmKt}) is divergent due to the
translational invariance of the background. Therefore, it is
convenient to consider the density ${\cal W}$: $\int d^2 x{\cal
W}=W$. The integration over $t$ can be easily performed. The
subsequent analytic continuation to $s=0$ yields a finite result,
\begin{equation}
{\cal W}_m =-\frac {m^3}{12\pi} \,.\label{cWm}
\end{equation}

By collecting the contributions from all elementary excitations on
this background we obtain:
\begin{equation}
{\cal W}^{\rm 1-loop} = -\frac{e^3}{6\pi} \left[ \left( 3|\phi |^2
-v^2 \right)^{3/2} -\left( 2|\phi |^2 \right)^{3/2} \right]
\,.\label{W1loop}
\end{equation}

Although the effective potential (\ref{W1loop}) is convergent
in $2+1$ dimensions, there are finite renormalization effects
which shift classical values of $e$ and $v$. To fix these shifts
we consider
\begin{equation}
{\cal W}^{\rm tot}={\cal W}^{\rm cl}(e+\hbar \delta e,v +\hbar\delta v)
+\hbar {\cal W}^{1-loop} \,,\label{Wtot} 
\end{equation}
where we have re-inserted the $\hbar$ dependence. The first term
on the right hand side is just the classical potential 
\begin{equation}
{\cal W}^{\rm cl}(e,v)=\frac{e^2}2 \left( |\phi |^2 -v^2 \right) 
\label{Wclass}
\end{equation}
with shifted values of $e$ and $v$. We require that to the first
order in $\hbar$ the potential ${\cal W}^{\rm tot}$ has a minimum
at $|\phi |=v$ (``no tadpole'' condition). This condition yields
\begin{equation}
\delta v = -\frac{e}{4\sqrt{2}\pi}\,.\label{deltav}
\end{equation}
To fix $\delta e$ one needs also another normalisation condition,
but for our purposes (\ref{deltav}) is already enough.

The shift (\ref{deltav}) induces a shift in the vacuum energy:
\begin{equation}
\Delta E^{\rm f.r.}=\hbar (\delta v)\frac{dE^{\rm cl}}{dv}=
-\frac{evn\hbar}{\sqrt{2}} \,.\label{DEfr}
\end{equation}
Since other contributions (\ref{1term}) and (\ref{DEBren}) vanish,
\begin{equation}
\Delta E^{\rm ren}=-\frac{evn\hbar}{\sqrt{2}} \,.\label{finres}
\end{equation}
This completes the calculation of the mass shift of the supersymmetric
vortex.
%%%%%%%%%%%%
\section{Conclusions}\label{conclu}
In this paper we have re-calculated one-loop quantum corrections
to the mass of the supersymmetric ANO vortex. We put the system
into a box with a circular boundary and applied the zeta function
regularization. We have demonstrated that boundaries violate
a part of the supersymmetries, but the remaining invariances are
enough to guarantee coincidence of the eigenfrequencies in the
bosonic and fermionic sectors. Therefore, contributions from
the bosons and the fermions to the vacuum energy cancel each 
other both in the full theory (vortex in a box) and in the 
effective theory near the boundary. Up to this point we agree with
the previous works \cite{Schmidt:cu,Lee:1994pm} (though our
conclusion is based on somewhat more reliable grounds). There is,
however,  a contribution (\ref{finres}) to the vacuum energy which comes from
finite renormalization of the couplings in the classical mass
of the vortex\footnote{This situation is similar to the BPS
black hole mass shift discussed in \cite{Rey:1996sm}.}. 
Such contribution was neglected in the approach
of \cite{Schmidt:cu,Lee:1994pm}\footnote{It was 
pointed out to
the present author by R.Wimmer that finite renormalizations 
will lead to a
nonvanishing correction.
}. To see what happens with the BPS
bound one has to calculate also quantum corrections to the central
charge.

Let us now give some comments on the vortex mass corrections in
a pure bosonic theory. These comments are motivated by the 
discussion \cite{discussion} on renormalization of the Casimir
energy. In the supersymmetric case  
it was essential, that we the bosonic and fermionic contributions
are cancelled mode-by-mode. In purely bosonic theory no such
cancellation may appear and the vacuum energy will be, in general,
divergent\footnote{In the zeta function regularization the one-loop
divergences are defined by the heat kernel coefficients. 
For the (mixed) boundary conditions used in this work the heat
kernel expansion can be found in \cite{Branson:1999jz}.}.
There are two types of the divergences which are given by volume
or by boundary integrals. Normally, boundary divergences are the
same in the full theory and in the effective theory defined near
the boundary when $R\to \infty$. Therefore, $(\Delta E(V+B) -
\Delta E(B))$ will contain volume divergences only which can
be removed by some standard renormalization procedure. However,
to define $\Delta E(V+B)$ or $\Delta E(B)$ separately one has
to introduce new surface counterterms which are absent in the
original model.  
%%%%%%%%%%%%
\section*{Acknowledgements}
I am grateful to M.~Bordag, A.~Rebhan and especially to R.~Wimmer
for fruitful discussions.
This work was supported in part by the DFG project BO 1112/12-1.
%%%%%%%%%%%
\appendix
%%%%%%%%%%
\section{Supertransformation and boundary conditions}\label{susyapp}
In this Appendix we show how one can prove compatibility
of our boundary conditions with the supersymmetry transformations
of the spinor fields with pure imaginary $\eta_-$.
Let us consider the supertransformation of $\chi_-$:
\begin{equation}
\delta \chi_-=-\eta_- \left( \partial_0 w +i\epsilon_{ojk}
\partial^j \alpha^k -2i e \Re ( \phi^* \varphi ) \right) \,.
\label{delchim}
\end{equation}
We are going to prove that $\delta \chi_-$ satisfies the same
boundary conditions as $\chi_-$ if $\Re (\eta_-)=0$. The condition
(\ref{bcU2V1}) on $U_2=\chi_-$ can be checked easily:
\begin{equation}
\Im (\delta \chi_-) |_{\partial M} \sim -\partial_0 w |_{\partial
M}=0 \,,\label{apa1}
\end{equation}
where we have used the boundary condition (\ref{bcw}). Let us now
check the boundary condition (\ref{bcUUN}):
\begin{equation}
0=\partial_r \Re (\delta \chi_-)|_{\partial M} \sim
\partial_r \left( -\epsilon_{ojk}
\partial^j \alpha^k + 2e \Re ( \phi^* \varphi ) \right)|_{\partial M}
\,.\label{apa2}
\end{equation}
Consider the term on the right hand side of (\ref{apa2}) which
contains $\alpha$:
\begin{eqnarray}
&& -\partial_r \epsilon_{ojk} \partial^j \alpha^k|_{\partial M} =
\frac 1r \left[ \left( \partial_r - \frac 1r \right) \partial_r
\alpha_\theta + \frac 1r \partial_\theta \alpha_r
-\partial_r\partial_\theta \alpha_r \right]|_{\partial M} \nonumber \\
&&\qquad\qquad =\frac 1r \left[ -(\Delta \alpha)_\theta -\left (
\partial_r + \frac 1r \right)
\partial_\theta \alpha_r -\frac 1{r^2} \partial_\theta^2 \alpha_\theta
\right]|_{\partial M}\nonumber \\
&&\qquad\qquad =-\frac 1r(\Delta \alpha)_\theta |_{\partial M}
\label{apa3}
\end{eqnarray}
where we first re-expressed the left hand side through $\alpha_r$
and $\alpha_\theta$, then we used the vector Laplacian in the
polar coordinates (cf., e.g., \cite{Vassilevich:we}):
\begin{eqnarray}
&&-(\Delta \alpha)_r = \left( \partial_r^2 + \frac 1r \partial_r +
\frac 1{r^2} \partial_\theta^2 -\frac 1{r^2} \right) \alpha_r -
\frac 2{r^3} \partial_\theta \alpha_\theta \,,\nonumber \\
&&-(\Delta \alpha)_\theta = \left( \partial_r^2 -\frac 1r
\partial_r +\frac 1{r^2} \partial_\theta^2 \right) \alpha_\theta +
\frac 2r \partial_\theta \alpha_r \,.\label{apa4}
\end{eqnarray}
Finally, to obtain the last line of (\ref{apa3}) we made use of
the boundary conditions (\ref{asbc}). The equations of motion for
$\alpha_\theta$ yield\footnote{More precisely, the equation to
follow is obtained by varying (\ref{L2bos}) with respect to
$\alpha_\theta$ and then using the Bogomolny equation (\ref{bog1})
for the background $\phi$.} :
\begin{equation}
-\frac 1r (\Delta \alpha)_\theta = -\frac 1r \omega^2
\alpha_\theta - 2e \left( \varphi^* \partial_r \phi +
\varphi\partial_r \phi^* \right) \,. \label{apa5}
\end{equation}
Now we collect all contributions to see
\begin{equation}
\partial_r \Re (\delta \chi_-) |_{\partial M}\sim
\left[ -\frac 1r \omega^2 \alpha_\theta + e\Re (\phi^*
\partial_r \varphi -\varphi \partial_r \phi^*) \right]|_{\partial M}=0
\label{apa6}
\end{equation}
due to (\ref{asbc}) and (\ref{bcphi}).

Calculations for other components of the spinor fields can be done
in a similar manner. 
%%%%%%%%%%

\end{document}